# The detection of non-Gaussian vibrations with improved spatial resolution and signal-to-noise ratio in distributed sensing

Qian He, Rong Liu, Chengdan Tan, Lijun Tang and Xiongjun Shang

*Abstract*—In fiber-optic distributed sensing, vibration signals are mostly assumed to follow Gaussian distribution for the simplicity of signal processing. However, in real applications, vibration signals often behave as non-Gaussian processes, which have rarely been highly considered. In this paper, a higher-order cumulants algorithm based phase-sensitive optical time-domain reflectometry (OTDR) is proposed to detect and analyze non-Gaussian vibration signals accompanied with noises. When disturbances are applied on the sensing fiber, the distribution probability of Rayleigh backscattering signals will deviate from the ideal Gaussian distribution. The non-Gaussian vibration is then extracted from Gaussian noises based on the probability density distribution. Simulations and experiments are carried out. The experimental results show that the demonstrated method can measure non-Gaussian vibrations with improved signal-to-noise ratio and spatial resolution.

## I. INTRODUCTION

Fiber-optic distributed sensing based on Rayleigh scattering is a powerful technique to detect the stress changes along the fiber. Phase-sensitive OTDR (φ-OTDR) attracts attentions due to its capability of high-sensitivity, distributed measurements and cost-effective [1]. The φ-OTDR is widely used in perimeter security, structure health monitoring and pipeline surveillance, etc [2] [3]. Reporting disturbance locations with high reliability and precision are vital in intelligent fault diagnoses. The exiting methods used to detect vibration signals are mostly based on the hypothesis that the distributions of vibration signals follow Gaussian distributions. However, in real applications, vibration often behaves as non-Gaussian and non-stable signals, e.g., the gear/rolling bearing works abnormally or an engine breakdown.

In previous researches, the method of moving differentials is applied to locate the vibration point [4]. The location signal-to-ratio (SNR) of ~6dB at the vibration point is obtained by analyzing 200 consecutive backscattering traces. Zhu etc. demonstrates reduced noise floor and enhanced the frequency response range [5] by use of moving average and moving differential methods, where 100 consecutive traces are processing to obtain SNR of ~6.7Db at fault locations. Later, a technique using continuous wavelet transform (CWT) is adopted in the φ-OTDR system to give the frequency-domain and time-domain information of the vibration event simultaneously [6]. The time-frequency information of the vibration is determined by the CWT scalogram with wavelet ridge detection, and the position information is acquired by the wavelet global spectrum, which is determined by the amplitude differences among the global wavelet spectrums of time sequences. A two-dimensional edge detection method has been proposed to extract location information of intruder in the φ-OTDR system [7]. The location SNR of 8.4 dB is achieved by two-dimensional edge detection using the Sobel operator. Spectral subtraction reducing wide-band background noise to enhance spectrum SNR of vibration signals is proposed [8]. A system based on frequency-division-multiplexing time-gated digital optical frequency domain reflectometry is proposed and realized the gains of spectrum SNR over 10 dB [9]. The distributed vibration sensing by sub-Nyquist additive random sampling was demonstrated recently [10], and achieved a broadband vibration sensing.

Even though existing techniques can efficiently remove noises from vibration signals, the exploitations for vibrations have been restricted by the hypothesis of Gaussian distributions. In real applications, dynamic signals often behave as non-Gaussian process, which have rarely been concerned in the previous researches of vibration sensing.

Statistical characteristics are available tools for the measurement of non-Gaussian signals. In this paper, we propose a higher-order cumulants (HOC) method based on φ-OTDR to detect non-Gaussian vibration signals with improved SNR and spatial resolution. The HOC algorithm can distinguish non-Gaussian signals and Gaussian noises by the statistics of probability density distribution [11]. First, the tendency of amplitude changes is subtracted from the average of backscattering traces; second, the HOC of backscattering traces are calculated to extract non-Gaussian features; finally, the vibration signals are located by value of HOC. The proposed technique can decrease Gaussian noises, e.g. the state of polarization (SOP) noise and the thermal noise. The proposed system can detect and locate non-Gaussian vibrations with high reliability and precision.

Qian He, Rong Liu, Heng Jiang and Xiongjun Shang are all with the Hunan Provincial Key Laboratory of Flexible Electronic Materials Genome Engineering, Changsha University of Science & Technology, Changsha 410114, China (Corresponding e-mail: hqian@csust.edu.cn)

## II. THEORY

Cumulants can reveal the statistic characteristic of time-domain signal. Given a sequence $K_n = \{x_{k1}, x_{k2}, ..., x_{kn}\}$, where $k$ denotes a collection of k stationary random variables with zero-mean. The function $f(x_1, ..., x_k)$ is the joint probability density of the sequence. The Fourier transform of $f(x_1, ..., x_k)$ can be expressed as

$$\Phi(\omega_1, \cdots, \omega_k) = E\{\exp j(\omega_1 x_1 + \cdots + \omega_k x_k)\} \\ = \int_{-\infty}^{+\infty} \cdots \int_{-\infty}^{+\infty} f(x_1, \cdots, x_k) e^{j(\omega_1 x_1 + \cdots + \omega_k x_k)} dx_1 \cdots dx_k, \quad (1)$$

where $\omega_i$ represent the angular frequency. The joint cumulants of order $r=r_1+...+r_k$ are defined as

$$c_{r_1, \cdots, r_k} \triangleq (j)^r \left. \frac{\partial^r \ln \Phi(\omega_1, ..., \omega_k)}{\partial \omega_1^{r_1} ... \partial \omega_k^{r_k}} \right|_{\omega_1 = \cdots = \omega_k = 0}, \quad (2)$$

where $r_i$ is the $i$ th-partial derivative of $\omega_i$. When $r_1=...=r_k=1$, we have

$$c_k = c_{r_1, \cdots, r_k} = cum(x_1, \cdots, x_k). \quad (3)$$

In the case of $x_i$ ($1 \leq i \leq k$) with non-zero-mean, $x_i$ are replaced by $x_i - E\{x_i\}$ in the above formulas. The simplified expressions of $c_3$ can be written as [12]

$$cum(x_1, x_2, x_3) = E\{x_1 x_2 x_3\} \quad (4)$$

The 1st-order cumulant represents the mean of variables and the 2nd-order cumulant is the variance of variables. The 3rd-order cumulant shows the degree of deviation of symmetry of $x$ compared to Gaussian distribution. The 3rd-order cumulant can suppress noises with symmetrical distributions. Therefore, the 3rd-order cumulant equals zero for the symmetrical Gaussian distribution. The expression is more complex with higher order. In this paper, the relationship between of asymmetrical of the probability density distribution and cumulant is studied, thus the 3rd-order cumulant is used.

For a stationary random time-sequence $x(t)$ with zero-mean, $c_{k,x}(\tau_1, \tau_2, ..., \tau_k)$ is denoted as the joint $k$ th-order cumulant of $x(t)$,

$$c_{k,x}(\tau_1, \tau_2, \cdots, \tau_k) = cum(x(t), x(t+\tau_1), \cdots, x(t+\tau_{k-1})). \quad (5)$$

The data estimation of the 3rd-order cumulant is approximated by

$$c_{3,x}(\tau_1, \tau_2) \simeq \frac{1}{N} \sum_{t=1}^{N} x(t) x(t+\tau_1) x(t+\tau_2), \quad (6)$$

where $\tau$ is the time lag between the selected and reference elements in the time sequence $x(t)$. Thus, the non-Gaussian signal can be extracted with high SNR from the accumulative Gaussian noises, which have the zero 3rd order HOC.

## III. SIMULATIONS

Detections and analysis of asymmetric non-Gaussian signals by HOC algorithm are simulated using MATLAB. $K_n$ and $c_{kn}$ ($n=1, 2, 3, 4, 5, 6$) are time sequences and the corresponding HOC for $k$ th sequences, respectively. The value of $K_1$ is the ideal time-sequence data, which follows the standard normal distribution with zero-mean and variance of one. We intercept the interval of $[-3.46, +3.46]$ as the domain of the distribution. The occupancy rate of remainder area enclosed by x-axis and function is 0.06% in the case of ideal Gaussian distribution.

In order to show the values of the 3rd order HOC of asymmetrical of the same data length variables. The probability density distribution functions of $K_2$ to $K_6$ are artificially asymmetrical based on the distribution of $K_1$. The fragmentary intervals are [-3.46, -1.28), [-1.28, -0.84), [-0.84, -0.52), [-052,-0.25), [-0.25, 0.00) for $K_2$ to $K_6$, respectively [Fig.1 (a)], which induce 10% asymmetry on the standard Gaussian distribution. The data length of $K_1$ is 99947, so that the maximum deviation among of those areas is 0.21% compared to the standard distribution.

We have $\tau_1=\tau_2=0$ when calculating HOC to improve the measurement performance of signals with low SNR and to avoid positioning error. The calculated $c_{kn}$ ($n=1, 2, 3, 4, 5, 6$) are shown in Fig.1(b). The $c_{k1}$ is 0 due to the symmetric distribution of $K_1$. The peak value of HOC appears at $K_2$ because of the strongest asymmetry of probability density distribution. The Gaussian signal is blinded by the HOC algorithm, thus the asymmetric non-Gaussian signal can be extracted effectively. The simulation results indicate that the value of HOC decreases as the symmetry of the distribution increases.

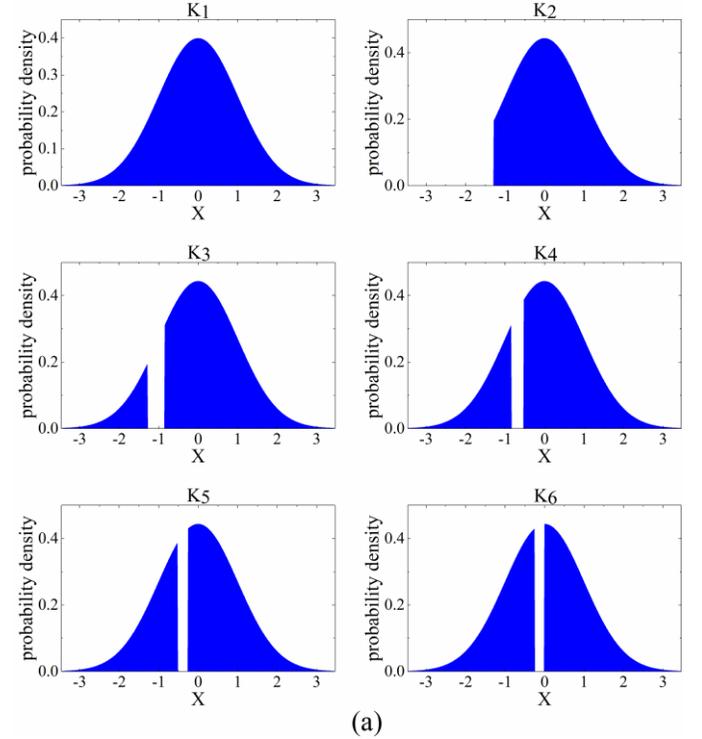

(a)

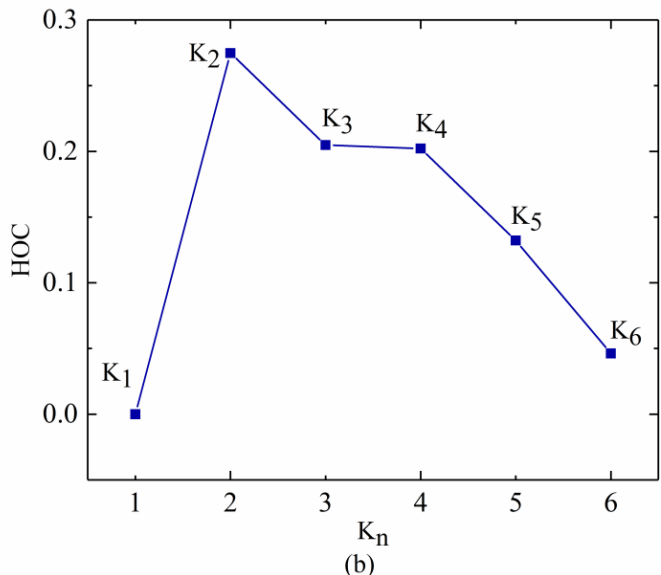

Fig. 1 (a) Probability density distributions of $K_1$ to $K_6$; (b) the corresponding value of the 3rd order HOC.

We define the calculated length as the number of sampling points used to calculate the 3rd order HOC. In order to show the effect of the calculated length of data sequence and the SNR of the original time-sequence ($SNR_1$) on the performance of the algorithm, the detections of squared wave signals with duty cycle of 10%, 20%, 30% and 40% are generated with Gaussian noise under different signal to noise ratio. Here the SNR=10 $\lg(P_{signal}/P_{noise})$, where $P_{signal}$ is the power of signal, and $P_{noise}$ is the average power of noises. The $SNR_1$ is the proportion of the power of the $P_s$ which is the squared wave signal to the power of the $P_n$ which is the power of the Gaussian noise in the mixed-signal. We define $SNR_2$ as the result the ratio of a signal mixed the squared wave signal and the Gaussian noise ($s+n$) to a signal that the pure Gaussian noise ($n'$) in same power, that $P_{(s+n)}=P_{(n')}$. The ratio of ($P_{signal}/P_{noise}$) in $SNR_2$ is the ratio of the mixed-signal's HOC value to the pure noise's HOC value, so the ratio is $(HOC_{s+n})/(HOC_{n'})$. Figure 2(a) gives simulation results of the SNR ($SNR_2$) with various calculated lengths ($SNR_1$=0). The result clearly shows that increasing calculated length improves the SNR of the time-sequence after calculated by the HOC algorithm and also enhance the stability of the algorithm. Figure 2(b) gives the relationship between $SNR_2$ and $SNR_1$ at computing length of 120. The results indicate that the HOC algorithm can improve the SNR efficiently.

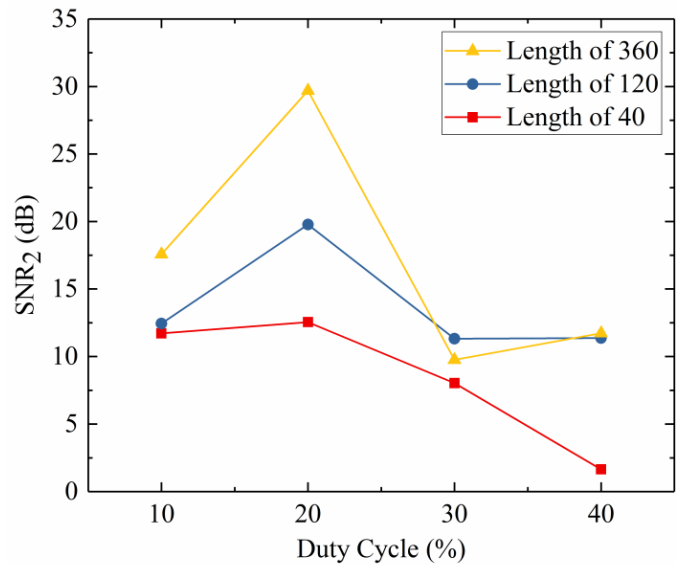

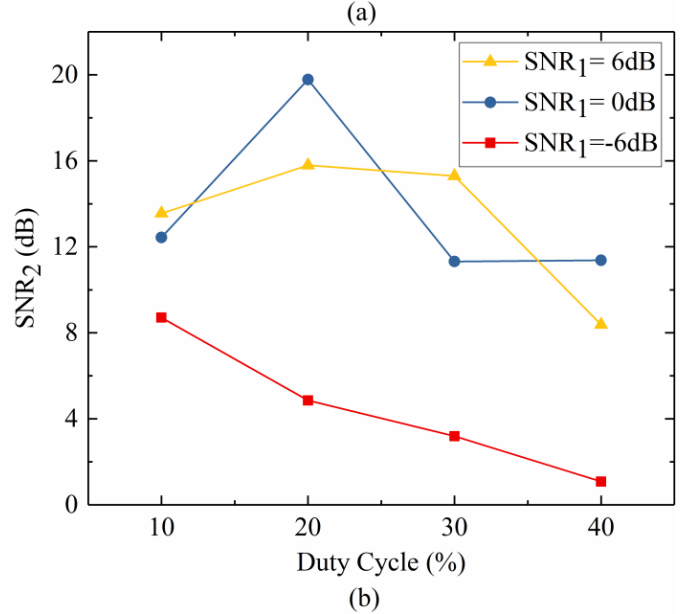

**Fig. 2** The effect of the computing length and the SNR of the original time-sequence on the performance of the algorithm (a) $SNR_2$ with various computing lengths ($SNR_1$=0); (b) $SNR_2$ with various $SNR_1$ at computing length of 120.

IV. EXPERIMENTAL RESULTS AND DISCUSSIONS

The schematic diagram is illustrated in Fig.3. A narrow line-width laser (NLL) with line-width less than 100 kHz outputs continuous laser. The average optical power is 10 mW. The light source is then modulated into optical pulses by an acoustic-optic modulator (AOM), which is driven by a waveform generator (WG). After amplified by an erbium-doped fiber amplifier (EDFA), the light pulses are injected into the sensing fiber through an optical circulator (OC). The Rayleigh backscattering light travels back through the fibre under test (FUT) and detected by a photoelectric detector (PD), and then sampled by a data acquisition card (DAQ) with sampling rate of 100 MS/s.

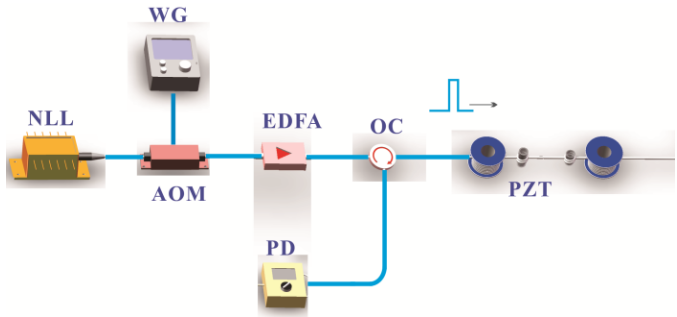

Fig. 3. The schematic diagram of phase-OTDR based distributed sensing system.

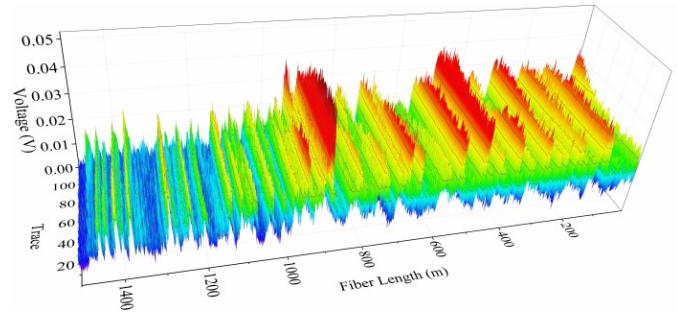

Fig. 5. 3-D image of original backscattering traces.

The value of HOC at each sampling point along the sensing fiber is calculated by 100 consecutive backscattering traces. As shown in Fig. 4, $x_n(1 \leq n \leq 1500)$ is the sampling points along the sensing fiber.

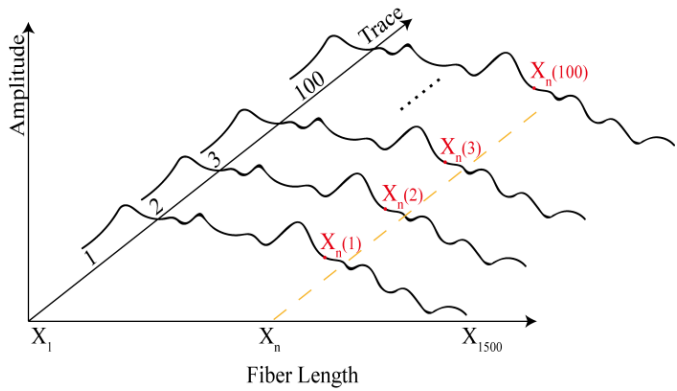

Fig. 4. The sampling Rayleigh backscattering traces.

Considering Eq.6, the HOC of $x_n$ can be written as

$$c_{xn} = \frac{1}{100}\sum_{i=1}^{100}[x_n(i)]^3. \quad (7)$$

The Rayleigh backscattering light can reflect the physical features of the sensing fiber because of the interferences within the optical pulse. The amplitudes probability distribution of Rayleigh backscattering mixed with white noise behave as Gaussian distribution. When disturbances are applied on the sensing fiber, the amplitude probability of the Rayleigh backscattering signals will deviate from the Gaussian distribution. The deviation value can be calculated to locate the vibration.

The amplitudes of interference light are sensitive to the random change of the state of polarization (SOP) of the interference light. Considering the central limit theorem, the probability distribution of SOP noise is close to the standard Gaussian distribution with sufficient sequence length. Therefore, the proposed system can also reduce the SOP noise

A Piezo-electric transducer (PZT) cylinder with ~1 m length of sensing fiber wounded on is used as a vibration source, which is placed at ~1049m location of the sensing fiber with a total length of 1500 m. The PZT was driven by electrical square wave signals with frequency of 700 Hz and duty cycles of 10%, 20%, 30% and 40%, and the electrical amplitude is 0V- 2V.

The probe pulses are injected into the sensing fiber with repetition rate of 10 kHz and pulse width of 100 ns. The sampled backscattering traces are presented in Fig.5. The vibrations are located by the HOC algorithm, and the experimental results are shown in Fig. 6. A clear peak value corresponding to the vibration point can be observed. The small deviation from the symmetrical distribution can lead to an obvious increase in the value of HOC.

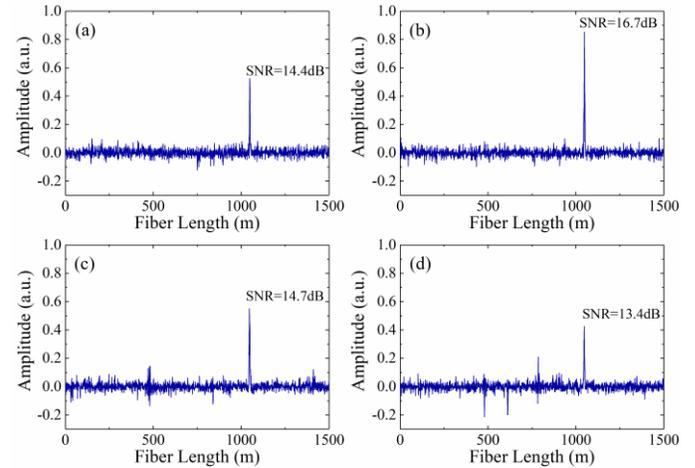

Fig. 6. Location information of vibrations.(a)vibration signal with duty cycle of 10%; (b)vibration signal with duty cycle of 20%; (c)vibration signal with duty cycle of 30%; (d)vibration signal with duty cycle of 40%.

The SNR of location signals is illustrated in Fig. 7. The peak value of 3rd-cumulant appears at vibration signal with duty cycle of 20%. As the power of vibration signals is submerged by noises, the original SNR of the vibration signal is difficult to measure. Therefore, we estimate the original SNR of the vibration signal is between 0 to 6dB based on the simulation result [Fig.2 (b)]. The experimental results clearly indicate that the proposed method can enhance SNR of vibration signal effectively.

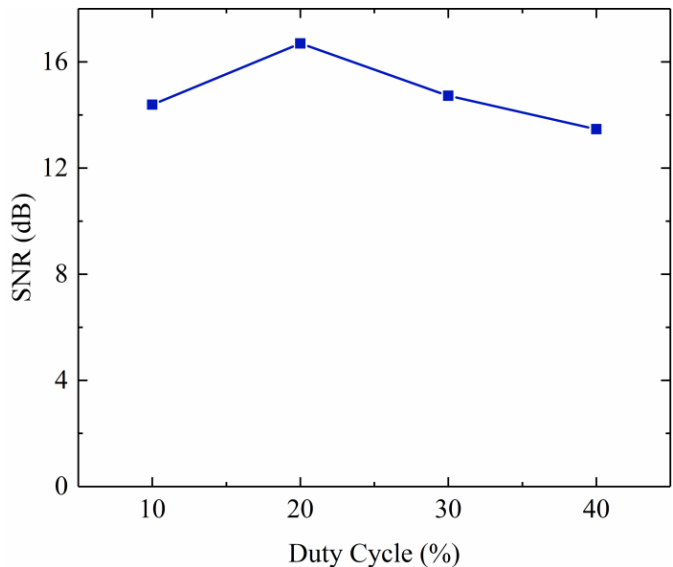

Fig. 7  The SNR of location signals.

The zoomed-in position profile of the vibration point is shown in Figure 8.

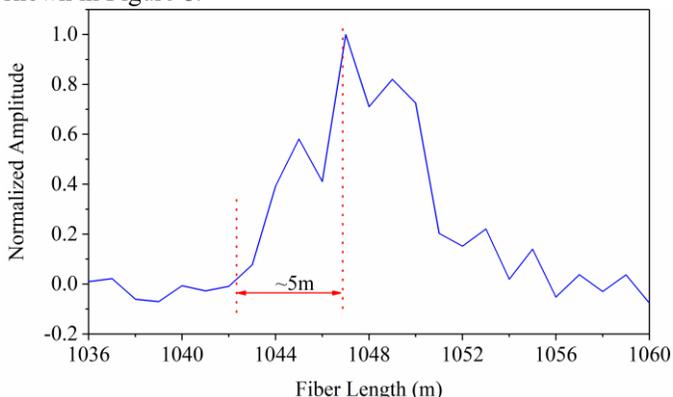

Fig. 8  Spatial resolution of the sensing system.

The spatial resolution (SR) is defined as the corresponding distance from 10% to 90% of the rising edge. Theoretically, the spatial resolution is 10m corresponding to 100ns pulse width. As shown in Fig.8, the SR is improved to ~5 m, which shows that the SR is improved by using the HOC algorithm.

The backscattering amplitudes change when the optical pulses pass through the vibration point. It can be considered reasonably that the internal stress change caused by vibrations is stable during the propagation of optical pulses when the vibration frequency is lower than ~10MHz. Figure 9 (a) illustrates two consecutive optical pulses travel through the vibration position at $t_1$ and $t_2$, respectively. The interference length $w_2$ within the optical pulses increases with the pulse transmitting through the vibration point. The amplitude changes at vibration point accumulate with the increasing interference length, and reach the peak value when the vibration point is in the middle of the optical pulse. The change induced by vibrations for each sensing point within the optical pulse is identical when the pulse is transmitting through the vibration point. Therefore, the rate of change for each sensing point at $t_1$ and $t_2$ is the same, resulting in linear difference between the amplitudes at the vibration point, as the dashed green line presented in Fig. 9 (b). The rate of amplitude change is nonlinear when the amplitude is processing by the 3rd-cumulant algorithm, which is a cubic operation. Thus, the rising edge and falling edge are shortened efficiently. Therefore, the HOC algorithm can further improve the SR.

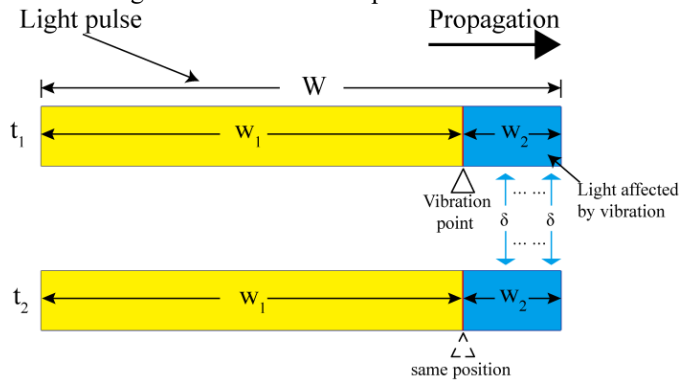

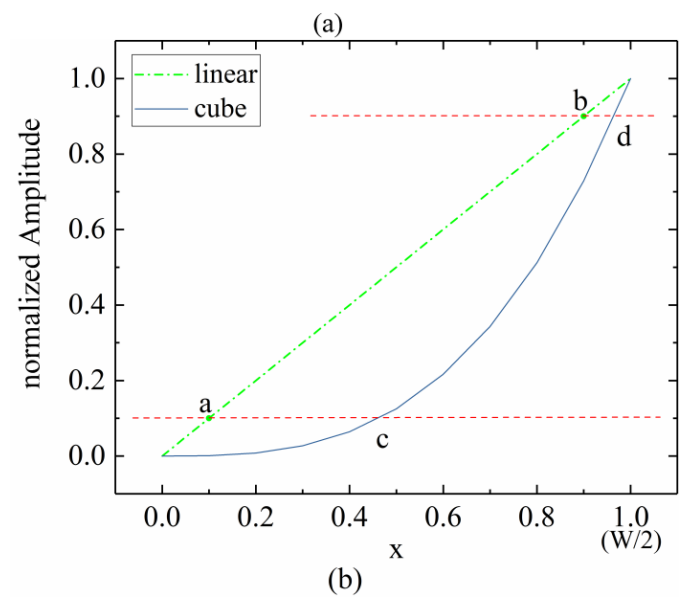

Fig.9. Theoretical analysis of the enhanced spatial resolution.

Moreover, the HOC analysis, including the value of HOC and the bi-spectrum, can be used to obtain and recognize signal patterns by showing the phase coupling among harmonics of signal frequencies, which is a promising technique in intelligent faults monitoring.

V.  CONCLUSION

In conclusion, a higher-order cumulants based phase-sensitive OTDR to detect the non-Gaussian vibration is presented in this paper. Improved SNR and spatial resolution support that the HOC algorithm is a useful tool to detect and analysis non-Gaussian vibration signals.